\documentclass[letterpaper]{twocolartcl}
\pdfoutput=1
\newcommand{\ct}{c_{\textrm{T}}}

\newcommand{\vc}{v_{\textrm{c}}} 



\title{\texorpdfstring{\vspace{-3.0em}\normalfont}{}%
       Limiting velocities and transonic dislocations in Mg}

\author{Khanh Dang, Daniel N. Blaschke, Saryu Fensin, Darby J. Luscher}

\date{}

\graphicspath{{./}{./figures/}}

\begin{document}

 \maketitle

\begin{strip}
\vspace{-2.2cm}
\begin{center}
\large
Los Alamos National Laboratory, Los Alamos, NM, 87545, USA
\\\ttfamily{\normalsize E-mail: kqdang@lanl.gov, dblaschke@lanl.gov, saryuj@lanl.gov, djl@lanl.gov}
\end{center}
\vspace{-0.4cm}
\hspace{0.02\textwidth}\parbox{0.96\textwidth}{
\begin{abstract}
To accurately predict the mechanical response of materials, especially at high strain rates, it is important to account for dislocation velocities in these regimes. 
Under these extreme conditions, it has been hypothesized that dislocations can move faster than the speed of sound. 
However, the presence of such dislocations remains elusive due to challenges associated with measuring these experimentally. 
In this work, molecular dynamics simulations were used to investigate the dislocation velocities for the basal edge, basal screw, prismatic edge, and prismatic screw dislocations in Mg in the sub-, trans-, and supersonic regimes. 
Our results show that only prismatic edge dislocations achieve supersonic velocities.
Furthermore, this work demonstrates that the discrepancy between the theoretical limiting velocity and the MD results for Mg is due to its sensitivity to large hydrostatic stress around the dislocation core, which was not the case for fcc metals such as Cu.
\end{abstract}
}
\end{strip}


\section{Introduction and motivation}
\label{sec:intro}
The highest achievable strain rates during high rate plastic deformation in metals are determined not only by the mobile dislocation density but also by the limiting velocities of dislocations via Orowan's relation \cite{Hansen:2013,Luscher:2016,Blaschke:2019a,Blaschke:2021impact}.
The latter can be calculated from linear elasticity theory in the limit of perfect, steady-state dislocations, while neglecting details of the dislocation core \cite{Blaschke:2021vcrit,Teutonico:1961,Teutonico:1962}.
Theory in this case predicts a diverging dislocation self-energy at these limiting (or critical) velocities $\vc$, which further depends on the dislocation character, crystallographic structure, and slip systems.
In the isotropic limit, all $\vc=\ct$, the transverse sound speed.
The steady-state theory of perfect dislocations also predicts the existence of transonic and supersonic dislocations, which are separated from the subsonic ones by critical velocity ``barriers'' \cite{Rosakis:2001,Weertman:1980,Weertman:1961}.
These limiting velocities subsequently lead to a divergence in the dislocation drag coefficient, which accounts for dislocation motion being impeded by phonon scattering \cite{Blaschke:2018anis,Blaschke:2019Bpap} --- even for accelerating dislocations \cite{Blaschke:2020acc}.
It has been argued, however, that these velocity barriers can in principle be overcome when the dislocation core is taken into account in a regularizing fashion \cite{Markenscoff:2008,Pellegrini:2018,Pellegrini:2020}, and hence should not be viewed as rigid limits, but rather as barriers in dislocation velocity that require extreme conditions to overcome \cite{Blaschke:2020MD}.

In some metals, this latter view is supported by a body of molecular dynamics simulations \cite{Olmsted:2005,Marian:2006,Daphalapurkar:2014,Tsuzuki:2008,Tsuzuki:2009,Oren:2017,Ruestes:2015,Gumbsch:1999,Li:2002,Jin:2008,Peng:2019,Blaschke:2020MD}.
Most notably, pure edge dislocations in some fcc metals seem to asymptotically approach the lowest shear wave speed (which in this special case happens to coincide with $\vc$) until some critical stress above which transonic motion becomes possible \cite{Olmsted:2005,Marian:2006,Daphalapurkar:2014,Tsuzuki:2008,Tsuzuki:2009,Oren:2017}.
Similar results were found for bcc tungsten whose second order elastic constants are ``almost isotropic'', in the sense that $c_{11} \approx c_{12}+2c_{44}$, in \cite{Gumbsch:1999,Li:2002,Jin:2008}.
Supersonic pure screw dislocations in fcc metals have only been found in Cu at very low temperatures, but not at room temperature \cite{Blaschke:2020MD,Peng:2019,Oren:2017,Olmsted:2005,Marian:2006}.

As for non-cubic metals, some MD simulations have also been conducted for hcp Mg, in particular for pure screw and pure edge dislocations in the basal, prismatic, and pyramidal slip systems.
It was found that with increasing shear stress, those dislocations approach a (subsonic) velocity \cite{Groh:2009} that is significantly lower than the theoretically expected limiting velocity \cite{,Blaschke:2021vcrit}.
Importantly, since the applied stress considered in \cite{Groh:2009} is less than 200 MPa, the transonic regime of dislocation mobility for hcp Mg remains unexplored.

Here, the mobilities of basal and prismatic dislocations in hcp Mg are more thoroughly investigated using molecular dynamics (MD) and density functional theory (DFT) simulations.
Supersonic dislocation motion is observed for prismatic edge dislocations in Mg.
In addition, we discover an explanation for the discrepancy between the theoretical limiting velocity and the MD results.
In particular, we show that the elastic constants and thus limiting velocity of Mg are significantly more sensitive to hydrostatic pressure than those of Cu.

\section{Limiting velocities in hcp basal and prismatic slip systems}
\label{sec:theory}

For ideal, straight dislocations (i.e. neglecting the dislocation core), theory predicts limiting velocities which follow from solving the differential equations
\begin{align}
	\partial_i \sigma_{ij}  &= \rho \ddot{u}_j
	\,, &
	\sigma_{ij} &= C'_{ijkl} \partial_l u_{k}
	\label{eq:diffeqns1}
\end{align}
in coordinates aligned with the dislocations.
This means that the $Z$ direction is aligned with the dislocation line and the $Y$ direction is parallel to the slip plane normal.
Thus, $C'_{ijkl}$ denotes the components of the tensor of second order elastic constants (SOEC) after rotation into the present coordinate basis.
For details on how to derive $\vc$ for arbitrary slip systems and dislocation character angles, we refer the interested reader to the recent review article of Ref. \cite{Blaschke:2021vcrit}.
Here, we merely quote the results for pure screw and edge dislocations in the hcp basal and prismatic slip systems with Burgers vector directions $\langle \bar2110 \rangle$.
All four cases feature a reflection symmetry discussed in detail in Ref. \cite{Blaschke:2021vcrit} (i.e. the following rotated elastic constants vanish: $c'_{14}$, $c'_{15}$, $c'_{24}$, $c'_{25}$, $c'_{46}$, $c'_{56}$), enabling the derivation of analytic expressions for $\vc$.
Thus, for pure screw dislocations, the limiting velocity is
\begin{align}
	v_c^\text{screw} &= \sqrt{\frac{1}{\rho}\left(c'_{55}-\frac{(c'_{45})^2}{c'_{44}}\right)}
	\,. \label{eq:vcscrew}
\end{align}
where the rotated elastic constants for basal slip are given by
\begin{align}
	c'_{44} & = c_{44}\,, &
	c'_{45} & = 0\,, &
	c'_{55} & = (c_{11}-c_{12})/2
	\,,
\end{align}
and for prismatic slip,
\begin{align}
	c'_{44} & = (c_{11}-c_{12})/2\,, &
	c'_{45} & = 0\,, &
	c'_{55} & = c_{44}
	\,.
\end{align}

For pure edge dislocations, there are in general two cases to consider (see \cite{Blaschke:2021vcrit} for details), but the metal discussed in this paper (Mg) belongs to the simpler case where the limiting velocity can be expressed as
\begin{align}
	\vc^\txt{edge}=\sqrt{\frac{c'_{66}}{\rho}}
	\,, \label{eq:vcedge_special1}
\end{align}
and where the required rotated elastic constant is given by $c'_{66}=c_{44}$ for basal slip and by $c'_{66}=(c_{11}-c_{12})/2$ for prismatic slip.

\begin{table*}[h!t!b]
\centering
\caption{\label{tab:SOECvcrit}List of material densities, elastic constants relevant for the $\vc$, and theoretical limiting velocities $\vc$ of Mg at ambient conditions (i.e. room temperature and zero pressure) for the following four cases:
basal slip and pure screw (superscript `s,b') and pure edge (superscript `e,b'), as well as prismatic slip and pure screw (superscript `s,pr') and pure edge (superscript `e,pr').
All cases are listed using experimental data from Ref. \cite{CRCHandbook}, computed from the Pei EAM potential \cite{Pei:2018}, and computed from the Wu MEAM potential \cite{Wu:2015}.}
\begin{tabular}{lccc|cccc}
\hline
\hline
\vspace{0.05cm}
& $\rho$[g/ccm]	&	$\frac{c_{11} - c_{12}}2$[GPa]	& $c_{44}$[GPa]	& $\vc^\text{s,b}$[km/s]& $\vc^\text{e,b}$[km/s] & $\vc^\text{s,pr}$[km/s] & $\vc^\text{e,pr}$[km/s] \\
\hline 
Mg (exp)	& 1.74 & 16.69 & 16.35 & 3.097 & 3.065 & 3.065 & 3.097 \\
Mg (Pei)	& 1.7532 & 14.58 & 23.025 & 2.884 & 3.663 & 3.663 & 2.884 \\
Mg (Wu)	& 1.7741 & 19.4 & 18.0 & 3.307  & 3.185 & 3.185 & 3.307 \\
\hline
\hline
\end{tabular}
\end{table*}

Numerical results for these various cases are listed in Table \ref{tab:SOECvcrit}.
In particular, we give material densities, elastic constants relevant for the $\vc$, and theoretical limiting velocities $\vc$  for Mg at ambient conditions, as derived from experimental data obtained from the CRC handbook \cite{CRCHandbook} as well as derived from the embedded atom model (EAM) potential of Ref. \cite{Pei:2018} (henceforth denoted by `Pei potential') and the modified embedded atom model (MEAM) potential of Ref. \cite{Wu:2015} (henceforth denoted by `Wu potential').

Note that, in general, dislocations can exhibit up to three limiting velocities and we denote here by $\vc$ always the lowest limiting velocity.
Thus, by `transonic' we refer to velocities above $\vc$ but below the highest of the limiting velocities.
By `supersonic' we refer to velocities higher than all limiting velocities (which need not necessarily coincide with sound speeds of sound waves propagating in the same direction as the gliding dislocation, see Ref. \cite[sec. 3]{Blaschke:2021vcrit}).

\section{A molecular dynamics study of hcp Mg}
\label{sec:MDresults}

\begin{figure}[!htb]
\centering
\includegraphics[width=0.75\columnwidth]{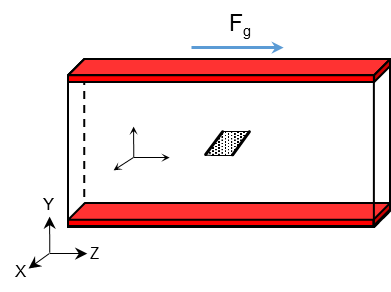}
\caption{Schematic of the simulation setup where a straight dislocation is generated via the Volterra displacement field \cite{Dang:2019,Dang:2020}.}
\label{fig:simbox}
\end{figure}

\begin{figure*}[!htb]
\centering
\includegraphics[width=0.75\textwidth,trim=1.pt 1.pt 1.pt 0, clip]{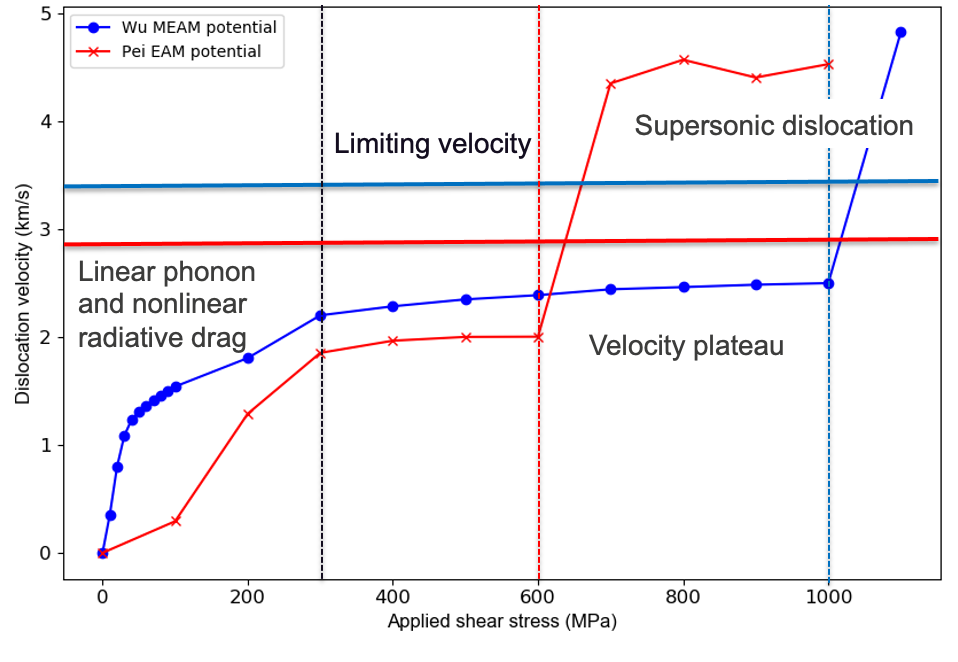}
\caption{Mobility curves for prismatic edge dislocations in Mg using Wu \cite{Wu:2015} and Pei \cite{Pei:2018} interatomic potentials at 100 K.
Both potentials exhibit transonic (or as we will show shortly, in fact supersonic) motion above some critical stress and after `plateauing' at a velocity which is lower than initially expected (solid line).
The two potentials also disagree on what the critical stress is.
We explain these discrepancies in the main text.}
\label{fig:plateaus}
\end{figure*}

\begin{figure*}[htb]
\centering
\includegraphics[width=1.\textwidth,trim=2.pt 1.pt 5.pt 1.pt, clip]{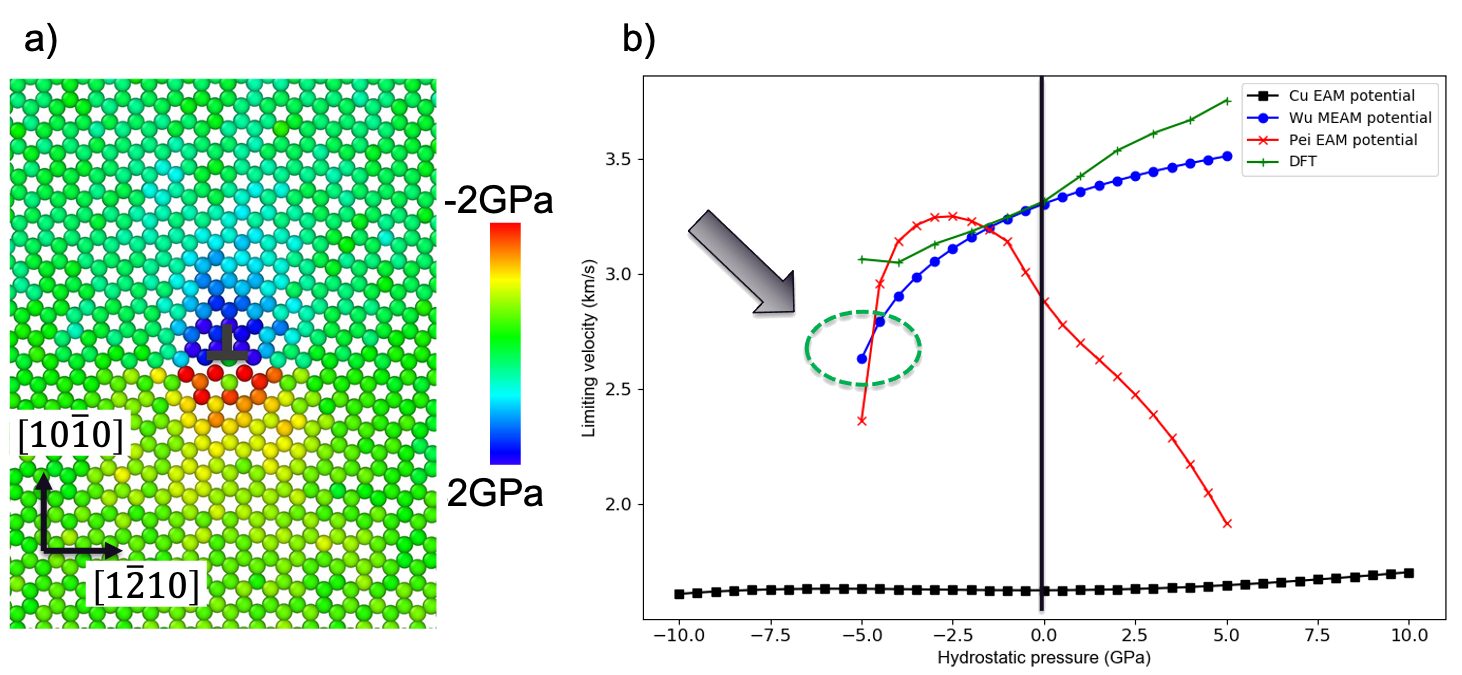}
\caption{a) Hydrostatic stress around a prismatic edge dislocation under 100 MPa shear stress.
The atoms are colored by hydrostatic stress, where positive values indicate compression and negative values indicate tension.
The range of pressures shown in the color bar is chosen to be $(-2,2)$ GPa to better highlight the contrast between negative and positive pressures.
b) The limiting velocity as a function of hydrostatic pressure using MD simulations with both Mg potentials, a Cu EAM potential as well as DFT calculations for Mg.
The limiting velocity is very sensitive to pressure in Mg, more so than in copper.
Since the two MD potentials yield very different results, we include a DFT calculation (green curve) which indicates that the Wu MEAM potential \cite{Wu:2015} is more reliable for our present purpose.}
\label{fig:elasticconstants}
\end{figure*}

\begin{figure*}[htb]
\centering
\includegraphics[width=1.0\textwidth,height=0.408\textwidth,trim=15.pt 4.pt 0.pt 7.pt, clip]{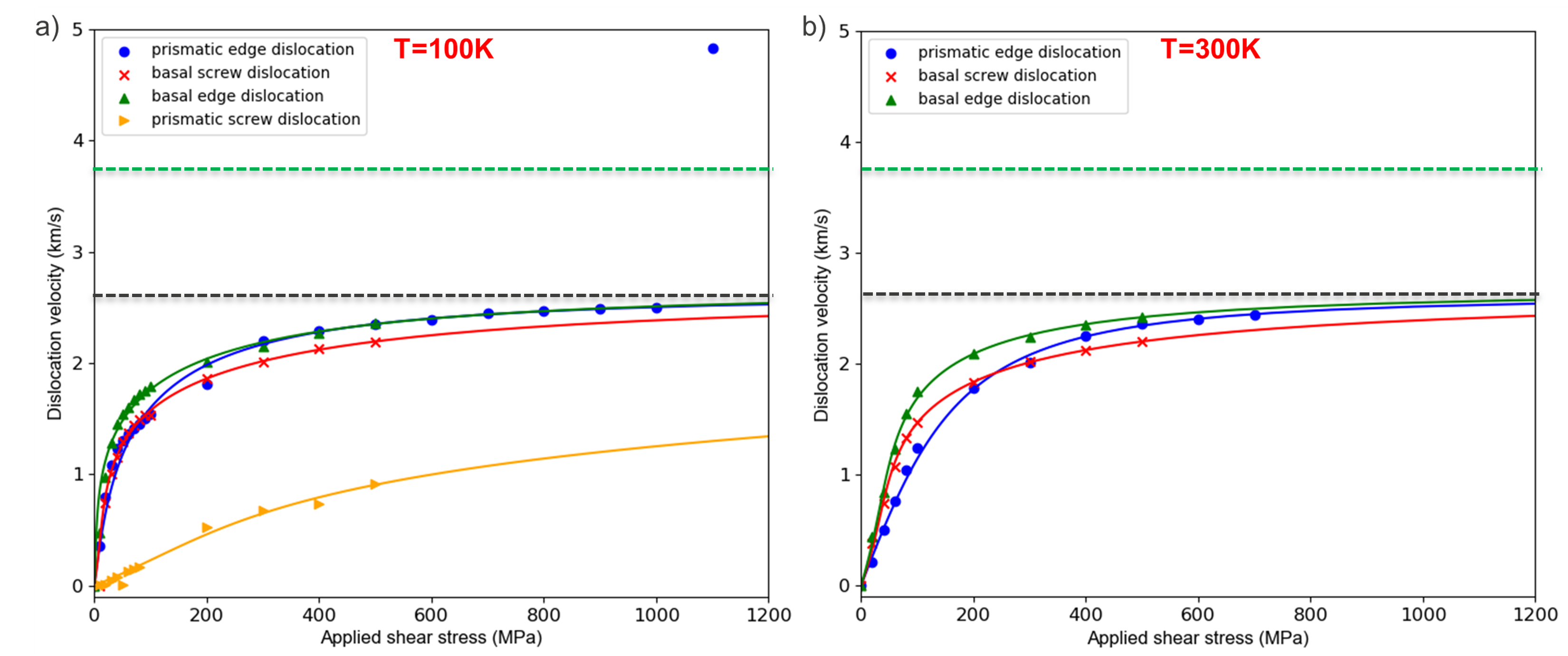}
\caption{Mobility curves for specified dislocations in Mg using Wu \cite{Wu:2015} at a) 100  and b) 300K.
The data points are from MD simulations, while the solid lines are fitted functions using Eq. \eqref{eq:fits}.
Note that this equation cannot capture the jump to supersonic speeds observed in the simulation data for prismatic edge dislocations.
Fitted parameters for each dislocation are provided in Table \ref{tab:fittingparams}.
The black and green dashed lines are the lowest values for the lowest and highest  limiting velocity for hydrostatic pressure from -5 to 5 GPa from the MD simulation.
Since the edge and screw limiting velocities for both temperatures and slip systems are relatively close (i.e. in the range 2.63--2.67 km/s for the lowest and 3.79--3.91 km/s for the highest limiting velocity), only one is shown here for each the lowest and the highest limiting velocity.}
\label{fig:mobilityMg}
\end{figure*}


The mobility of dislocations is calculated using molecular dynamics simulation code LAMMPS \cite{Plimpton:1995} and tracked via the Open Visualization Tool (OVITO) with its Dislocation Extraction Algorithm (DXA) feature \cite{Stukowski:2009,Stukowski:2012}.
To properly capture the dislocation core structure and mobility, two recently developed interatomic potentials which are Wu et al. MEAM \cite{Wu:2015} and Pei et al. EAM \cite{Pei:2018} are used.
Both potentials can reproduce not only fundamental properties such as lattice parameters, cohesive energies, and elastic constants but also defect properties such as stacking fault energies and dislocation core structures, which are important for the study of dislocations and their mobility.

For this study, straight prismatic and basal dislocations with different characters (edge and screw) are inserted to the simulation cell as shown in Fig. \ref{fig:simbox}.
This was done using the Volterra displacement field with proper treatment of the boundary atoms as demonstrated by previous studies for mobility of dislocations in Al \cite{Dang:2019,Dang:2020}.
The prismatic slip system is chosen as $[1\bar{2}10]$ $(10\bar{1}0)$, while the studied basal slip system is $[1\bar{2}10]$ $(0001)$.
The $X$ and $Z$ directions coincide with dislocation line and gliding directions, respectively, while a mixed boundary condition is applied in the $Y$ direction.
The simulation cell is about $7.0 \times 19.0 \times 62.5$ nm in $X$, $Y$, and $Z$ directions, and contains approximately 385,000 atoms.
This size has been shown to be sufficient to ensure the dislocation mobility results  do not strongly depend on the image force effects from the free surface and the periodic boundary \cite{Dang:2019}.
Preliminary MD simulations of basal screw dislocations further confirm that this simulation box size has minor effects on the results.
The introduction of the dislocation is followed by an energy minimization routine to obtain relaxed dislocation core structures, which cannot be described by the Volterra displacement field.
The system with the relaxed dislocation configuration is then brought to 100K via a Nos{\'e}-Hoover style thermostat and barostat \cite{Melchionna:1993}.
Since the $Y$ direction is non-periodic and its pressure cannot be controlled via the barostat, the $Y$ position of the top boundary is manually adjusted using an iterative technique to relieve the pressure in the $Y$ direction caused by lattice thermal expansion at 100 K.
To calculate the elastic constants, the standard approach in LAMMPS is used where a small strain is applied in the appropriate directions and the change in stresses of the relaxed initial and strained configurations are calculated.
A similar approach is used in VASP for the DFT calculations \cite{Yu:2010,Kresse:1996a}.

Figure \ref{fig:plateaus} shows the mobility curve for edge dislocations moving on prismatic slip planes for both interatomic potentials.
Using the elastic constants derived from those potentials under ambient conditions, the limiting velocity was calculated according to \eqref{eq:vcedge_special1} with $c'_{66}=(c_{11}-c_{12})/2$ for prismatic slip, which is 3.3 and 2.9 km/s for Wu and Pei potentials, respectively.
Overall, each mobility curve contains three distinct regimes.
Below a driving stress of about 300 MPa, dislocation velocities strongly depend on stress due to the effects of phonon and radiative drag \cite{Dang:2019,Olmsted:2005}.
Above approximately 300 MPa, plateaus in  dislocation velocity are observed.
This is initially expected to occur close to the limiting velocity (solid line), as was the case in earlier studies of fcc metals \cite{Olmsted:2005,Marian:2006,Daphalapurkar:2014}.
However, in hcp Mg the dislocation velocity saturates at a lower value of stress, a result that is consistent with the earlier study of Ref. \cite{Groh:2009}.
Importantly, both Wu and Pei potentials show that transonic motion is possible, though the critical stress values for these ``jumps'' to occur differ significantly.

This leads to two critical questions regarding these results:
\begin{enumerate}
\itemsep0pt
\item[(1)]  Why does the dislocation velocity in hcp Mg behave differently compared to fcc metals by saturating at a lower value (limiting velocity) as compared to theoretical predictions?
\item[(2)] Which one of the two potentials is more reliable?
\end{enumerate}
Both questions can be addressed by exploring the effects of pressure (where a positive value refers to compression and a negative value is tension) on the elastic constants and eventually the limiting velocity.
From the dislocation theory point of view, the elastic fields near the core structure diverge to infinity.
With atomistic simulations, one can capture the finite values of stress associated with these fields.
Figure \ref{fig:elasticconstants}a) shows a typical hydrostatic stress distribution around the core of an edge dislocation in Mg.
The pressure ranges from $-4.6$ to $4.2$ GPa, where positive values indicate compression and negative values indicate tension.
Therefore, changes in the elastic constants were calculated associated with a hydrostatic stress ranging from -5 to 5 GPa using the two potentials for Mg. Additionally as a baseline, similar calculations were also performed for Cu.
In order to further validate the reliability of the two Mg potentials, MD simulation results for elastic constants were also compared to our calculated DFT values \cite{Yu:2010,Kresse:1996a}.
Figure \ref{fig:elasticconstants}b) shows how the lowest limiting velocity calculated from those elastic constants changes with pressure.
The limiting velocity of Cu does not change significantly as a function of pressure (even in the extreme range of $-10$ or $10$ GPa).
In contrast to Cu, the limiting velocity of Mg is strongly dependent on the hydrostatic pressure.

Thus, reasonable agreement with the plateau observed in MD simulations is achieved by arguing that the region of the dislocation field which, due to pressure (ranging from -4.6 to 4.2 GPa), has the lowest limiting velocity prevents the entire dislocation from moving any faster.
While both Mg potentials show strong dependence of limiting velocity on hydrostatic pressure, the results from the Wu potential (blue curve) agrees better with the DFT calculations, thus revealing that the Wu potential is more appropriate for the study of dislocation dynamics under extreme loading conditions.

Correcting elastic constant $c_{11}$ for pressure in the same way, we find that the largest of the three limiting velocities for pure edge dislocations decreases from roughly 6 km/s down to about 3.8 km/s, thus revealing that the highest dislocation velocity observed in our MD simulations for prismatic edge dislocations is truly supersonic (rather than transonic), see Figs. \ref{fig:mobilityMg} and \ref{fig:plateaus}.

\begin{table*}[h!t!b]
\centering
\begin{tabular}{l|c|c|c|c}
\hline
\hline
\vspace{0.05cm}
& $C_0$[$\mu$Pa-s]	&	$C_1$[$\mu$Pa-s]	& $C_2$[$\mu$Pa-s]& $C_3$[$\mu$Pa-s] \\
\hline 
Basal Edge & 3.05 / 19.1 & 9.27 / 21.7 & 0 / 0 & 62.7 / 45.4\\
Basal Screw & 10.1 / 23.3 & 22.78 / 36.8 & 2.98 / 2.56 & 86.3 / 84.2\\
Prismatic Edge & 8.4 / 2.43 & 0 / 0 & 1.31 / 1.14 & 40.6 / 29.1\\
Prismatic Screw & 152.1 & 214.7 & 0 & 1513\\
\hline
\hline
\end{tabular}
\caption{\label{tab:fittingparams}Fitted parameters for the mobility curves of all considered dislocations.
The first value is for 100K, the second one is for 300K.}
\end{table*}

\begin{table*}[h!t!b]
\centering
\begin{tabular}{l|c|c|c}
\hline
\hline
\vspace{0.05cm}
& $B(v=0.5\text{km/s})$ [$\mu$Pa-s]	&	$B({v=1\text{km/s}})$ [$\mu$Pa-s] & $B(v=1.5\text{km/s})$ [$\mu$Pa-s] \\
\hline 
Basal Edge & 7.69 / 49.9 & 14.3 / 45.7& 35.0 / 51.8\\
Basal Screw & 23.6 / 56.5 & 28.7 / 52.6 & 56.2 / 70.3\\
Prismatic Edge & 28.9 / 78.1 & 37.6 / 84.5 & 56.7 / 98.6\\
Prismatic Screw & 437 & 602 & 1107\\
\hline
\hline
\end{tabular}
\caption{\label{tab:dragcoeff}Drag coefficient at the different velocities.
The first value is for 100K, the second one is for 300K.}
\end{table*}

Figure \ref{fig:mobilityMg} shows the mobility curve for basal and prismatic dislocations with edge and screw character for the Wu interatomic potential at 100K.
The term `dislocation mobility' refers to the inverse of drag coefficient $B$.
Its defining equation reads $\sigma b=vB$, where $\sigma$ denotes the resolved shear stress driving the dislocation motion, $b$ is the Burgers vector length, and $v$ denotes the dislocation velocity \cite{Alshits:1992,Blaschke:2019Bpap,Blaschke:2018anis}.
As such, dislocation mobility ($1/B$) is directly proportional to the slope of the curves shown in Figure \ref{fig:mobilityMg}.
The regime of low driving stress is usually referred to as the `viscous regime' because, reminiscent of a friction constant, $B$ exhibits very little change initially.
As stress increases, $B$ transitions into a highly non-linear regime which is dominated by the dislocation's limiting velocity and therefore sometimes is called the `relativistic regime' \cite{Gurrutxaga:2020}.
As shown in \cite{Blaschke:2019fits}, dislocation drag $B$ can be quite accurately fit across the viscous and relativistic (but strictly subsonic) regime to the following functional form:
\begin{align}
B(v,\vartheta)&\approx
C_0(\vartheta) - C_1(\vartheta) x + C_2(\vartheta)\left(\frac1{\sqrt{1-x^2}} - 1\right) \nonumber\\
&\quad + C_3(\vartheta)\left(\frac1{(1-x^2)^{3/2}} - 1\right)
\,, \nonumber\\
x&=\frac{v}{\vc(\vartheta)}
\,, \label{eq:fits}
\end{align}
where $\vartheta$ denotes the dislocation character angle (0 for pure screw and $\pi/2$ for edge dislocations), $C_i$ are fitting parameters described in detail in Ref. \cite{Blaschke:2019fits}, and $\vc$ is the character dependent limiting velocity discussed 
above in Section \ref{sec:theory}.
Note that $B(v=0)=C_0$.
After an initial (roughly linear) drop parameterized by $C_1$, the drag coefficient $B(v)$ increases its value and is quickly dominated by the leading (subleading) divergence parametrized by $C_3$ ($C_2$) as the dislocation velocity approaches the limiting velocity $\vc$; see \cite{Blaschke:2018anis,Blaschke:2019fits} for details.

Table \ref{tab:fittingparams} shows fitting parameters $C_i$ for the subsonic regimes of the mobility curves shown in Figure \ref{fig:mobilityMg}
and Table \ref{tab:dragcoeff} compares the drag coefficient of all considered dislocation types at different velocities.
Among all considered dislocations, the prismatic screw dislocation has the lowest mobility, while the basal edge dislocation has the highest mobility in the low driving stress regime.
These trends are consistent with the Peierls stress calculations  reported for the Wu MEAM potential \cite{Wu:2015}.
These results are also in reasonable agreement with results from a previous study by Groh et al. \cite{Groh:2009} using the EAM Mg potential developed by Sun et al. \cite{Sun:2006}.
Specifically, the basal dislocations are more mobile than the prismatic dislocations. 
The main difference between results from this work (using the Wu et al. potential \cite{Wu:2015}) and the one from Groh et al. \cite{Groh:2009} is the mobility of the basal screw dislocation, where their results show that the basal screw dislocation is much less mobile compared to the basal edge and prismatic edge dislocations.
Experimentally, the Peierls stress of basal edge and screw dislocations have been reported to be relatively similar \cite{Tonda:2002,Stricker:2020}.
This is in reasonable agreement with the mobility curves obtained in this work.

Moreover, the  phonon drag (and thus high speed mobility) of dislocations reduces as temperature increases for all considered dislocations.
The supersonic speed is only observed at the lower temperature (100 K) similar to the previous study in Cu of Ref. \cite{Blaschke:2020MD}.
Importantly, unlike the prismatic edge dislocation, no transonic nor supersonic speed is observed for other dislocation types and characteristics.
Specifically, for basal edge and screw dislocations subjected to applied shear stress greater than 500 MPa, additional neighboring basal dislocations are formed and eventually creating long basal stacking faults (I$_2$) propagating throughout the simulation cell.
Even though the unstable stacking fault energy $(\gamma_\txt{us})$ for I$_2$ is relatively large (about 90--100 mJ/m$^{-2}$), the stable stacking fault energy is relatively small (only 22 mJ/m$^{-2}$) \cite{Wu:2015}.
This indicates that in high loading regimes, once the basal faults are formed, they will propagate and remain long.
On the other hand, the prismatic screw dislocation becomes unstable and transforms to the basal screw dislocation when subjected to applied shear stress greater than 500 MPa.
This is in agreement with previous results by Stricker and Curtin \cite{Stricker:2020} where they showed that (1) the prismatic screw dislocation is not stable and would transform into the dissociated basal screw dislocation and that (2) the prismatic slip occurs by a cross-slip process of this dissociated basal screw dislocation.
The stable core structure and the gliding motion of the prismatic screw dislocation predicted by the MEAM potential for applied stress less than 500 MPa is likely to be an artifact.
Finally, the prismatic edge dislocation has a stable core structure and there is no stable and lower energy stacking fault on the prismatic plane.
Therefore, the prismatic edge dislocation can remain stable at high applied stress and even cross over to the transonic and supersonic regimes.

\section{Conclusion}

In conclusion, the mobilities of edge and screw dislocations in Mg are studied with a focus on extreme loading conditions.
Only prismatic edge dislocations achieve supersonic velocity at lower temperature (100K) while the other three dislocation types either cross-slip or form long stacking faults for applied shear stress above 500 MPa.
It is also shown that the discrepancy between the theoretical limiting velocity and the MD results for Mg is due to its sensitivity to large hydrostatic stress around the dislocation core, which was not the case for fcc metals such as Cu.

\paragraph{Acknowledgements}
\ \\\small
The authors would like to thank M. Baskes and N. P. Daphalapurkar for related discussions.
We also thank the anonymous referees for their valuable comments.
Research presented in this article was supported by the Laboratory Directed Research and Development program of Los Alamos National Laboratory under project number 20210826ER.
This research used resources provided by the Los Alamos National Laboratory Institutional Computing Program, which is supported by the U.S. Department of Energy National Nuclear Security Administration under Contract No. 89233218CNA000001.
The authors also acknowledge support from the Materials project within the Physics and Engineering Models (PEM) Subprogram element of the Advanced Simulation and Computing (ASC) Program as well as the ASC Advanced Technology Development and Mitigation (ATDM) Subprogram element at Los Alamos National Laboratory (LANL) in the final stages of this work.

\bibliographystyle{utphys-custom-twocol}
\bibliography{dislocations}

\end{document}